\documentclass[aps,prc,showpacs, showkeys,floatfix,footinbib,twocolumn]{revtex4}

\usepackage{amsmath}    
\usepackage{graphicx}   
\usepackage{verbatim}
\usepackage{color}      
\usepackage{hyperref}   

\usepackage{mathptmx}
\long\def\/*#1*/{}
\newcommand*{\revision}{\textcolor{black}}

\RequirePackage{lineno}

\begin{document}

% \linenumbers %for revtex4-1

% \title{The van der Waals type interacting Hadron Resonance Gas model and the critical point in the $T$ versus $\mu_B$ phase diagram}
\title{Criticality in a Hadron Resonance Gas model with the van der Waals interaction}

\author{Subhasis Samanta}
\email{subhasis.samant@gmail.com}

\author{Bedangadas Mohanty}
\email{bedanga@niser.ac.in}

\affiliation{School of Physical Sciences, National Institute of Science 
Education and Research, HBNI, Jatni - 752050, India}

\begin{abstract}
The van der Waals interaction is implemented in a Hadron Resonance Gas model.
It is shown that this model can describe Lattice QCD data of
different thermodynamical quantities
satisfactorily with the van der Waals parameters $a = 1250 \pm 150$ MeV 
fm$^3$ and $r = 0.7 \pm 0.05$ fm. 
Further, a liquid-gas phase transition is observed in this model with the critical point
at temperature, $T = 62.1$ MeV and baryon chemical potential, $\mu_B = 708$ MeV.
\end{abstract}


\maketitle

\section{\label{sec:Intro} Introduction}
Lattice quantum chromo
dynamics (LQCD) ~\cite{Aoki:2006we,Borsanyi:2011sw,Bazavov:2012jq, Borsanyi:2013bia, Bazavov:2014pvz}
provides a first principle approach to study strongly interacting matter
at zero chemical potential ($\mu_B$) and finite temperature ($T$). 
LQCD calculations indicate a smooth cross over transition ~\cite{Aoki:2006we} from 
hadronic to a quark-gluon plasma (QGP) phase at zero baryon chemical potential 
and finite temperature \cite{Gupta:2011wh}.
On the other hand, at high 
baryon chemical potential and low temperature the nuclear matter is expected to have a
first-order phase transition ~\cite{Asakawa:1989bq} which ends
at a critical point, a second-order phase transition point as
one moves towards the high temperature and low baryon chemical potential region, 
in the QCD phase diagram \cite{Stephanov:1998dy,Stephanov:1999zu}. 
At present, the properties of QCD matter at high temperature
and small baryon chemical potential are being investigated using ultra relativistic 
heavy ion collisions at the Large Hadron Collider (LHC), CERN
and Relativistic Heavy Ion Collider (RHIC), Brookhaven National Laboratory (BNL).
The Beam Energy Scan (BES) program of RHIC \cite{Abelev:2009bw} is currently
investigating the location of the critical point \cite{Adamczyk:2013dal}.
The HADES experiment
at GSI, Darmstadt is investigating a medium with very large baryon chemical potential
\cite{Agakishiev:2015bwu}.
In future, the Compressed Baryonic Matter (CBM) experiment \cite{Ablyazimov:2017guv}
at the Facility for Antiproton and
Ion Research (FAIR) at GSI and the Nuclotron-based Ion Collider 
fAcility (NICA) \cite{Kekelidze:2017ual} at JINR, 
Dubna will also study nuclear matter at large baryon chemical potential.

The ideal or non-interacting Hadron Resonance Gas (HRG) model is quite successful in 
reproducing the zero chemical potential LQCD data 
of bulk properties of the QCD matter at low temperatures $T < 150$ MeV
~\cite{Borsanyi:2011sw, Bazavov:2012jq, Bazavov:2014pvz,Bellwied:2013cta, Bellwied:2015lba}.
However, disagreement between LQCD data and ideal HRG model calculations have been observed 
at higher temperatures. Considering excluded volume correction,
which mimics repulsive interaction, in HRG model, one can improve the picture in the crossover 
temperature region $T \sim $ 140-190 MeV 
~\cite{Andronic:2012ut, Bhattacharyya:2013oya}.
In the Excluded Volume HRG (EVHRG) model 
~\cite{Hagedorn:1980kb, Rischke:1991ke, Cleymans:1992jz, Yen:1997rv, Tiwari:2011km,
Begun:2012rf, Andronic:2012ut, Fu:2013gga, Tawfik:2013eua, Bhattacharyya:2013oya,Albright:2014gva,Vovchenko:2014pka,
Albright:2015uua,Alba:2016hwx,Kadam:2015xsa, Kadam:2015fza, 
 Kapusta:2016kpq}, 
effects of Van der Waals type hadronic repulsions at short distances are introduced but long distance 
repulsive interactions are ignored. Recently Van der Waals (VDW) type interaction with both
attractive and repulsive parts 
have been introduced
in HRG model ~\cite{Vovchenko:2015xja, Vovchenko:2015vxa, Vovchenko:2015pya, Redlich:2016dpb, Vovchenko:2016rkn}.
Interestingly VDWHRG model shows first order liquid-gas phase transition in nuclear 
matter at large chemical potentials and small temperatures which was not observed in other HRG models like
ideal HRG or EVHRG models. The liquid-gas phase transition in nuclear 
matter was also predicted in Refs. \cite{Sauer:1976zzf,Jaqaman:1983sg,Siemens:1984xiy}
and observed in experiment as well \cite{Karnaukhov:2003vp}.
In the Ref. ~\cite{Vovchenko:2015vxa} the VDW parameters $a$ and $b$ have been fixed by reproducing
the saturation density $n_0 = 0.16$ fm$^{−3}$ and binding energy $E/N = $ 16 MeV of
the ground state of nuclear matter. The
nuclear matter shows the critical point ~\cite{Vovchenko:2015vxa} at $T = 19.7$ GeV
and $\mu_B = 908$ MeV.
Latter it has been shown 
that using same interaction parameters $a$ and $b$ as for nuclear matter, for all the baryons in VDWHRG model LQCD data can be described 
qualitatively ~\cite{Vovchenko:2016rkn} in the cross over region.
The motivation of the present work is to
carry out the reverse prescription, that is to find out van der Waals parameters $a$ and $b$
that gives the best 
description of LQCD data
at zero chemical potential using VDWHRG model
and then extend this work to the finite chemical potential and
try to locate the existence of a critical point in the QCD phase diagram.

% The paper is organized as follows. First we describe the ideal HRG
% as well as VDWHRG model. Then we present our results.
% Finally we summarize our findings for this work.
The paper is organized as follows. In the Sec. \ref{sec:model} we describe the ideal HRG
as well as VDWHRG model. In Sec.\ref{sec:result} we present our results.
Finally in the Sec. \ref{sec:conclusions} we summarize our findings for this work.

\section{\label{sec:model} Model description}
There are varieties of HRG models which exis in the literature.
Different versions of this model and some of the recent works using these models
may be found in Refs. \cite{Hagedorn:1980kb, Rischke:1991ke,
Cleymans:1992jz, BraunMunzinger:1994xr, Cleymans:1996cd, Yen:1997rv, BraunMunzinger:1999qy,
Cleymans:1999st, BraunMunzinger:2001ip, BraunMunzinger:2003zd, Karsch:2003zq, Tawfik:2004sw,
Becattini:2005xt, Andronic:2005yp, Andronic:2008gu,Begun:2012rf, Andronic:2012ut,
Tiwari:2011km, Fu:2013gga, Tawfik:2013eua, Garg:2013ata, Bhattacharyya:2013oya,
Bhattacharyya:2015zka,Chatterjee:2013yga,Chatterjee:2014ysa,Chatterjee:2014lfa,Becattini:2012xb,Bugaev:2013sfa,Petran:2013lja,
Vovchenko:2014pka,Kadam:2015xsa, Kadam:2015fza, Albright:2014gva,
Albright:2015uua,Bhattacharyya:2015pra, Kapusta:2016kpq,Begun:2016cva,Adak:2016jtk, Xu:2016skm,Fu:2016baf,
Vovchenko:2015xja, Vovchenko:2015vxa, Vovchenko:2015pya,Broniowski:2015oha,Vovchenko:2015idt,
Redlich:2016dpb,Vovchenko:2016rkn,Alba:2016fku,Samanta:2017kmg,
Samanta:2017ohm,Sarkar:2017ijd,Bhattacharyya:2017gwt,Chatterjee:2017yhp,Alba:2016hwx}.
Some of the HRG models are non-interacting and some of them consider interaction 
among the particles. Next we will briefly discuss the non-interacting HRG model
and the HRG model with van der Waals type interaction.

% \subsection{Ideal HRG model}
In the ideal HRG model, the thermal system consists of non-interacting point like
hadrons and resonances.
The logarithm of the partition function of a hadron resonance gas 
in the grand canonical ensemble can be written as 
\begin {equation}
 \ln Z^{id}=\sum_i \ln Z_i^{id},
\end{equation}
where the sum is over all the hadrons and resonances. 
$id$ refers to ideal {\it i.e.}, non-interacting HRG model.
For particle species $i$,
\begin{equation}
 \ln Z_i^{id}=\pm \frac{Vg_i}{2\pi^2}\int_0^\infty p^2\,dp \ln[1\pm\exp(-(E_i-\mu_i)/T)],
\end{equation}
where $V$ is the volume of the thermal system, 
$g$ is the degeneracy, $E = \sqrt{p^2 + m^2}$
is the single particle energy, $m$ is the mass of the particle
and $\mu_i=B_i\mu_B+S_i\mu_S+Q_i\mu_Q$ is the
chemical potential. In the last expression, $B_i,S_i,Q_i$ are respectively
the baryon number, strangeness and electric charge of the particle, $\mu^,s$ are 
the corresponding chemical potentials.
The upper and lower sign of $\pm$ corresponds
to fermions and bosons, respectively.
We have incorporated all the hadrons and resonances listed in the particle 
data book up to a mass of 3 GeV \cite{Patrignani:2016xqp}.
The pressure $p^{id}$, the energy density 
$\varepsilon^{id}$ and the number density $n^{id}$ of the thermal system
are given by the following equations,

\begin{equation}
 p^{id} =\sum_i (\pm)\frac{g_iT}{2\pi^2}\int_0^\infty p^2\,dp \ln[1\pm\exp(-(E_i-\mu_i)/T)],
\end{equation}

\begin{align}\label{eq:e}
\begin{split}
\varepsilon^{id} =\sum_i \frac{g_i}{2\pi^2}\int_0^\infty\frac{p^2\,dp}{\exp[(E_i-\mu_i)/T]\pm1}E_i ,
\end{split}
\end{align}

\begin{equation}
 n^{id} = \sum_i \frac{g_i}{2\pi^2} \int_o^{\infty}  \frac{p^2 dp}{\exp[(E_i-\mu_i)/T \pm 1}.
\end{equation}
Once we know the partition function or the pressure of the system
we can calculate other thermodynamic quantities.

% \subsection{HRG model with van der Waals interaction}
The van der Waals equation in the canonical ensemble is given by
\cite{Book_griener}
\begin{equation}\label{eq:vanderWaals}
 \left(p + \left(\frac{N}{V}\right)^2 a \right) (V-Nb) = NT,
\end{equation}
where $p$ is the pressure of the system, $V$ is the volume, 
$T$ is the temperature, $N$ is the number of particles
and $a,b$ (both positive) are the van der Waals parameters. The parameters
$a$ and $b$ describe the attractive and repulsive interaction 
respectively.
The Eq. \ref{eq:vanderWaals} can be written as

\begin{equation}\label{eq:p}
 p(T,n) = \frac{NT}{V-bN} -a\left(\frac{N}{V}\right)^2 \equiv \frac{nT}{1-bn} - an^2,
\end{equation}
where $n \equiv N/V$ is the number density of particles.
The first term in the right hand side of the Eq. \ref{eq:p}
corresponds to the excluded volume correction where
the system volume is replaced by the available volume $V_{av} = V -bN$,
where $b = \frac{16}{3} \pi r^3$ is the proper volume of particles with $r$
being corresponding hard sphere radius of the particle.
The second term in Eq. \ref{eq:p} corresponds to the attractive interaction between particles.
The importance of van der Waals equation is that this analytical model
can describe first order liquid-gas phase transition of a real gas
which ends at the critical point. 
Such a feature is also an expectation for the QCD phase diagram.
% The critical point
% is characterized as
% \begin{equation}
%  \left(\frac{\partial p}{\partial n}\right)_T = 0, ~~ \left(\frac{\partial^2 p}{\partial n^2}\right)_T = 0. 
% \end{equation}
% The temperature, particle number density and pressure at critical point
% are given by \cite{Book_griener}
% \begin{equation}
%  T_c = \frac{8a}{27b}, ~~ n_c = \frac{1}{3b}, ~~ p_c = \frac{a}{27b^2}.
% \end{equation}

The van der Waals equation of state in the Grand canonical ensemble
can be written as \cite{Vovchenko:2015xja, Vovchenko:2015vxa}
\begin{equation}
 p(T,\mu) = p^{id}(T,\mu^*) - an^2, ~~ \mu^* = \mu - bp(T,\mu) -abn^2 + 2an,
\end{equation}
% where ideal gas pressure is given by
% \begin{equation}
%  p^{id}(T,\mu) =\pm\frac{gT}{2\pi^2}\int_0^\infty p^2\,dp \ln[1\pm\exp(-(E-\mu)/T)],
% \end{equation}
% where $g$ is the degeneracy of the particle, $E = \sqrt{p^2 + m^2}$
% is the single particle energy, $m$ is the mass of the particle.
% The upper and lower sign of $\pm$ corresponds
% to fermions and bosons respectively.
where $n \equiv n(T,\mu)$ is the particle number density of the van der Waals gas:
% The particle number density ($n$) and the entropy density ($s$) of the van der Waals gas can be written as
\begin{equation}
n \equiv n(T,\mu) \equiv \left(\frac{\partial p}{\partial \mu}\right)_T = \frac{n^{id}(T,\mu^*)}{1 + bn^{id}(T,\mu^*)}.
\end{equation}
% where particle number density for ideal gas is
% \begin{equation}
%  n^{id}(T,\mu) = \frac{g}{2\pi^2} \int_o^{\infty}  \frac{p^2 dp}{\exp[(E-\mu)/T \pm 1}.
% \end{equation}
The entropy density ($s$) for van der Waals gas can be written as
\begin{equation}\label{eq:s_vdw}
 s(T,\mu) \equiv \left(\frac{\partial p}{\partial T}\right)_{\mu} = \frac{s^{id}(T,\mu^*)}{1 + bn^{id}(T,\mu^*)}.
\end{equation}
Further, the energy density can be calculated as
\begin{equation}\label{eq:2ndlaw}
 \varepsilon (T,\mu) = T s + \mu n -p,
\end{equation}
and is given by
\begin{equation}
 \varepsilon (T,\mu) = \frac{ \varepsilon^{id} (T,\mu^*)}{1 + b n^{id}(T,\mu^*)} - a n^2.
\end{equation}
% For $a = b =0$ van der Waals equations are reduced to the ideal
% gas equations.
For a single component nuclear matter ($g =4, m =938$ MeV)
the values of van der Waals parameters were obtained as $a = 329$ MeV fm$^3$
and $b = 3.42$ fm$^3$ ($r = 0.59$ fm) \cite{Vovchenko:2015vxa} 
from the properties of the ground state of the nuclear matter.

For a hadronic system, 
we assume that interaction exist between all pair
of baryons and all pair of antibaryons.
We ignore the interaction for mesons in order
to avoid divergence of their number densities when modified
chemical potentials become close to the masses of the particles. 
The baryon-antibaryon interaction is also ignored because
the short range repulsive interaction between baryon and antibaryon
may be dominated by the annihilation processes \cite{Andronic:2012ut}.
These are the limitations of the current model and leaves the scope for further 
improvement in future.
Hence the pressure of VDWHRG model can be written as \cite{Vovchenko:2016rkn}
\begin{equation}
 p(T,\mu) = p_M(T,\mu) + p_B(T,\mu) + p_{\bar{B}}(T,\mu),
\end{equation}
with
\begin{equation}
 p_M(T,\mu) = \sum_{k \in M} p_k^{id}(T,\mu_k),
\end{equation}

\begin{equation}
 p_B(T,\mu) = \sum_{k \in B} p_k^{id}(T,\mu_k^{B*}) - a n^2_B,
\end{equation}
and
\begin{equation}
 p_{\bar B}(T,\mu) = \sum_{k \in {\bar B}} p_k^{id}(T,\mu_k^{\bar{B}*}) - a n^2_{\bar B},
\end{equation}
where $M, B, \bar{B}$ stand for mesons, baryons and antibaryons respectively.
The modified chemical potential for baryons and antibaryons
are given by
\begin{equation}
 \mu_k^{B (\bar{B})*} = \mu_k - b p_{B(\bar{B})} - ab n^2_{B(\bar{B})} + 2a n_{B(\bar{B})},
\end{equation}
where $n_B$ and $n_{\bar{B}}$ are particle number densities of
baryons and antibaryons respectively.
Once we know the pressure of the system, we can calculate different thermodynamic
quantities. The derivative of $p_{B(\bar{B})}$ with respect to the 
baryon chemical potential will give us the corresponding number densities:
\begin{equation}
 n_{B(\bar{B})} = \frac{\sum_{k \in B(\bar{B})} n_k^{id}(T,\mu_k^{B(\bar{B})*})}{1 + b \sum_{k \in B(\bar{B})} n_k^{id}(T,\mu_k^{B(\bar{B})*})}.
\end{equation}
From pressure, we can calculate entropy density, energy density using the Eqs. 
\ref{eq:s_vdw} - \ref{eq:2ndlaw}. Further one can calculate specific heat at
constant volume as 
\begin{equation}
C_V = \left( \frac{\partial{\varepsilon}}{\partial{T}} \right)_V
\end{equation}
and the  susceptibilities of conserved charges as
 \begin{equation}
  \chi_{BSQ}^{xyz} =\frac{\partial^{x+y+z} (p/T^4)}{\partial ({\mu_B/T})^x \partial ({\mu_S/T})^y \partial ({\mu_Q/T})^z}.
 \end{equation}
In the VDWHRG model if we put $a = 0$ and $b =0$ we will get the results of the ideal HRG model.
While with $a = 0$ in VDWHRG model it corresponds to Excluded Volume HRG (EVHRG) model
~\cite{Hagedorn:1980kb, Rischke:1991ke, Cleymans:1992jz, Yen:1997rv, Tiwari:2011km,
Begun:2012rf, Andronic:2012ut, Fu:2013gga, Tawfik:2013eua, Bhattacharyya:2013oya,Albright:2014gva,Vovchenko:2014pka,Albright:2015uua,Kadam:2015xsa, Kadam:2015fza, 
 Kapusta:2016kpq}, where only
repulsive interaction is included. Both ideal HRG model and EVHRG model
do not show any kind of phase transition.
Still these models are quite successful in describing LQCD data of the bulk properties of 
hadronic matter in thermal and chemical
equilibrium \cite{Borsanyi:2011sw, Bazavov:2012jq, Bazavov:2014pvz,Bellwied:2013cta, Bellwied:2015lba,
Karsch:2003zq, Tawfik:2004sw, Andronic:2012ut,Bhattacharyya:2013oya}.
This model is also successful in describing the ratios of hadron yields, 
at chemical freeze-out, created in central 
heavy ion collisions from SIS up to LHC energies~\cite{BraunMunzinger:1994xr, Cleymans:1996cd, 
BraunMunzinger:1999qy, Cleymans:1999st, BraunMunzinger:2001ip, Becattini:2005xt, 
Andronic:2005yp, Andronic:2008gu}. The heavy ion collisions at RHIC and LHC have
established quark-hadron phase transition.

\begin{figure*}[]
\centering
 \includegraphics[width=0.3\textwidth]{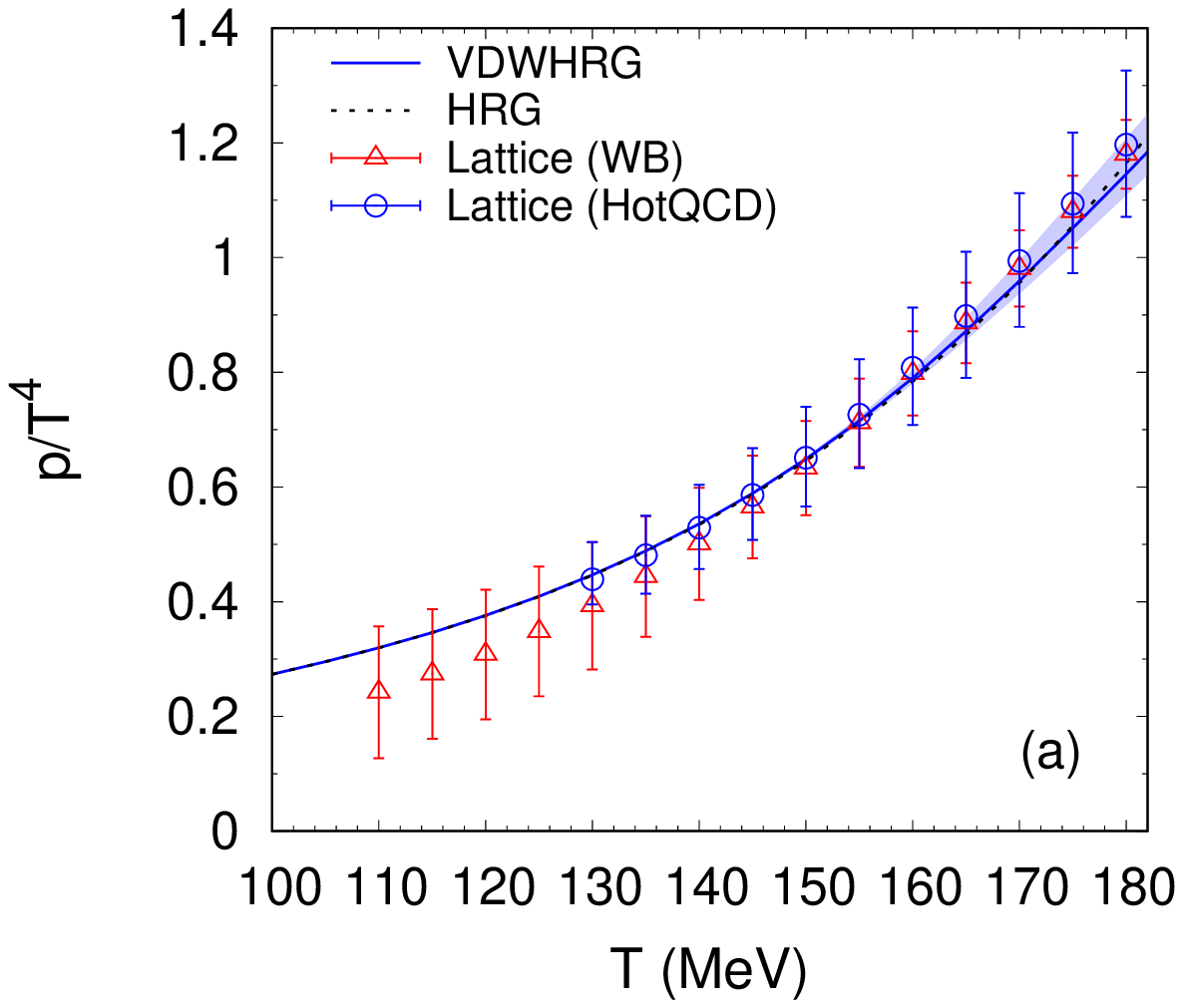}\label{PT4_T}
  \includegraphics[width=0.3\textwidth]{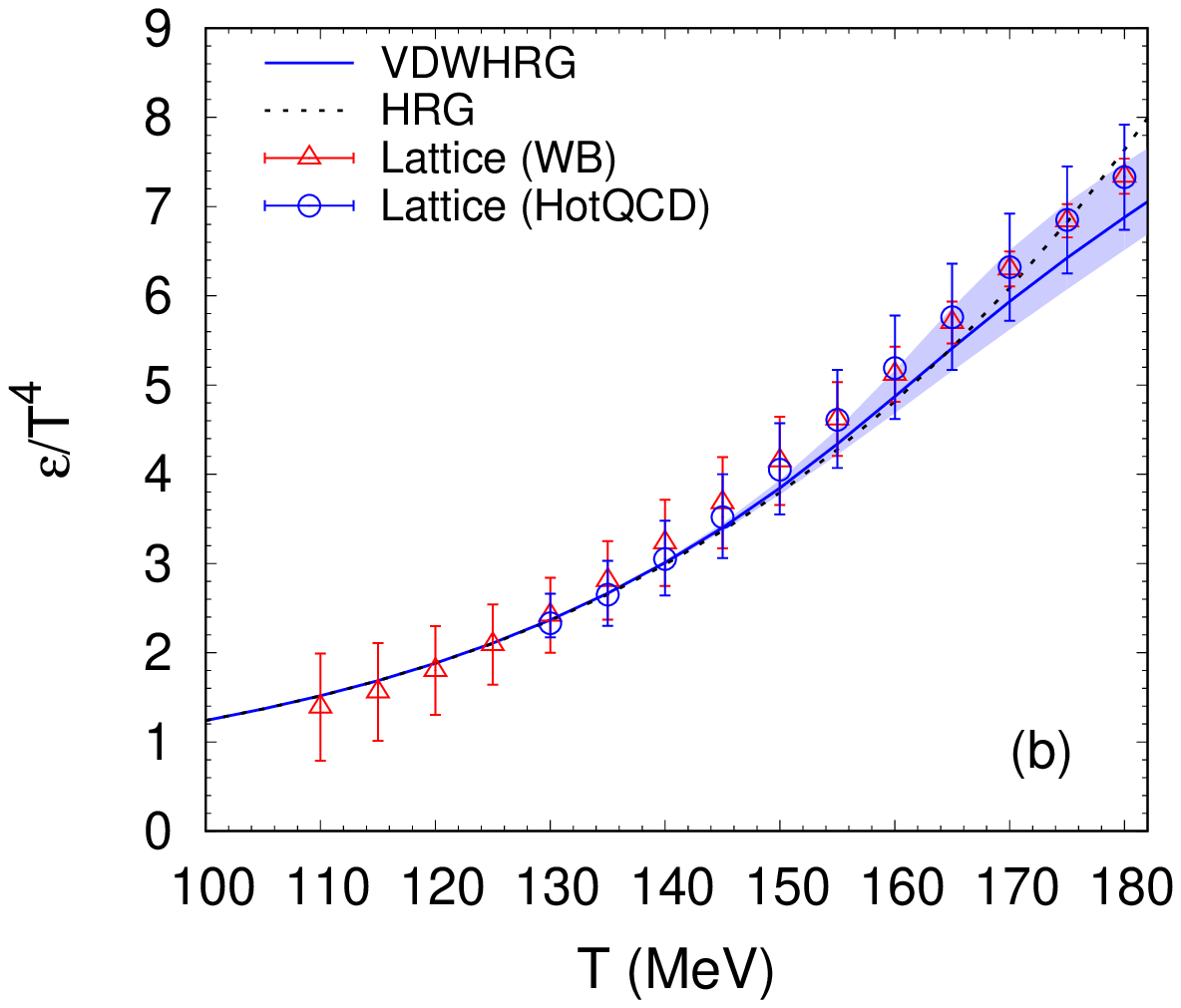}\label{ET4_T}
  \includegraphics[width=0.3\textwidth]{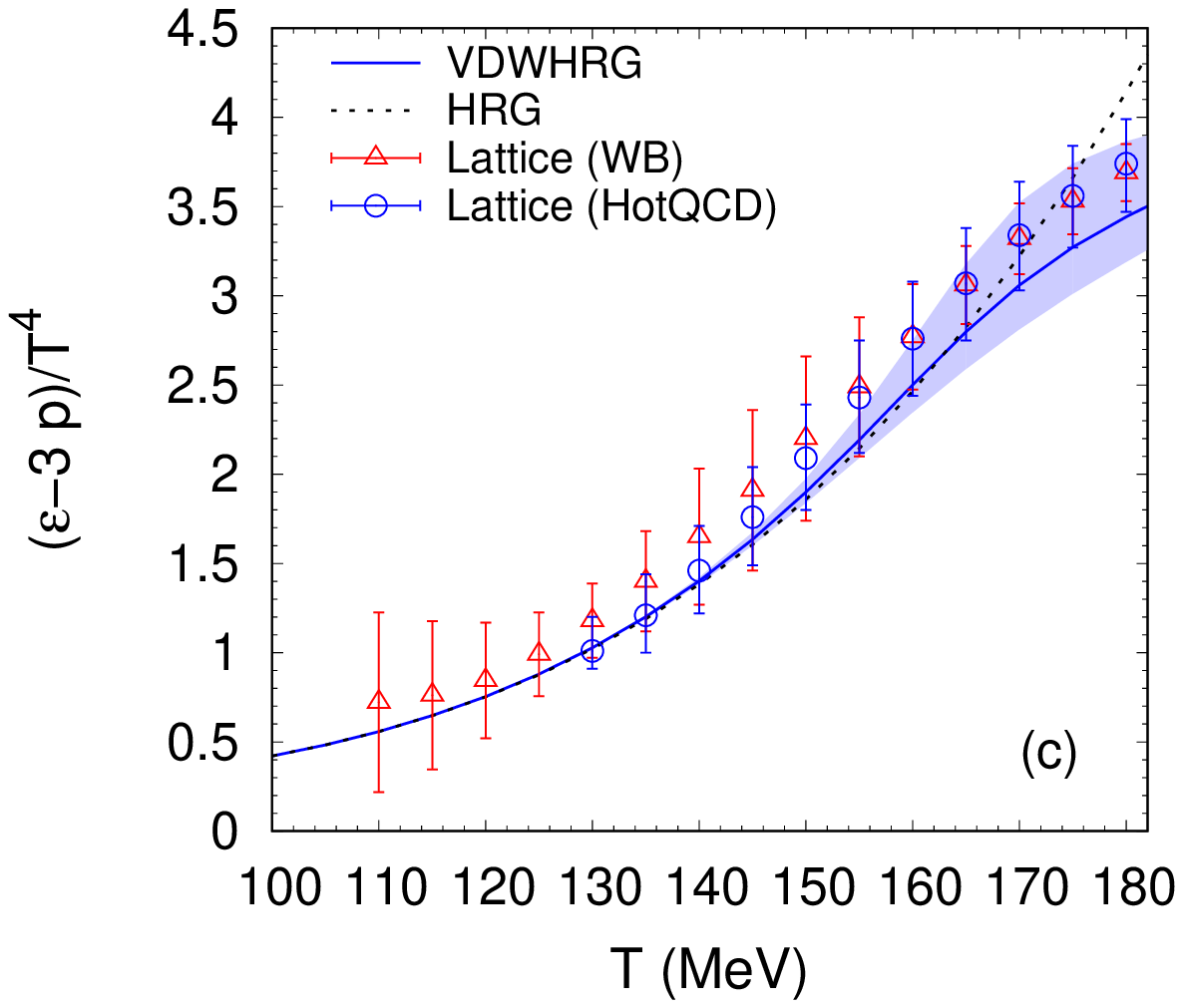}\label{tranom_T}
  \includegraphics[width=0.3\textwidth]{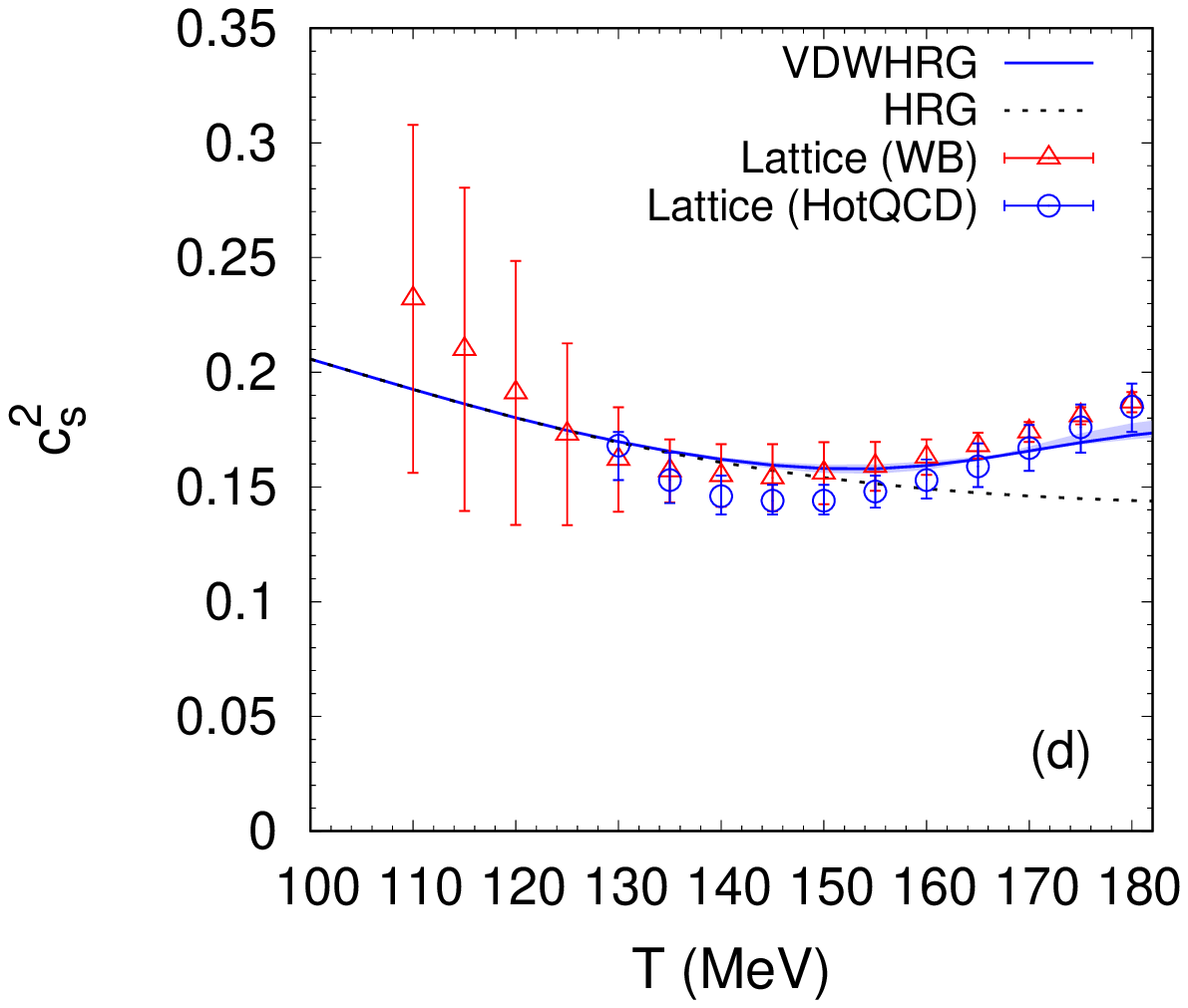}\label{cs2_T}
   \includegraphics[width=0.3\textwidth]{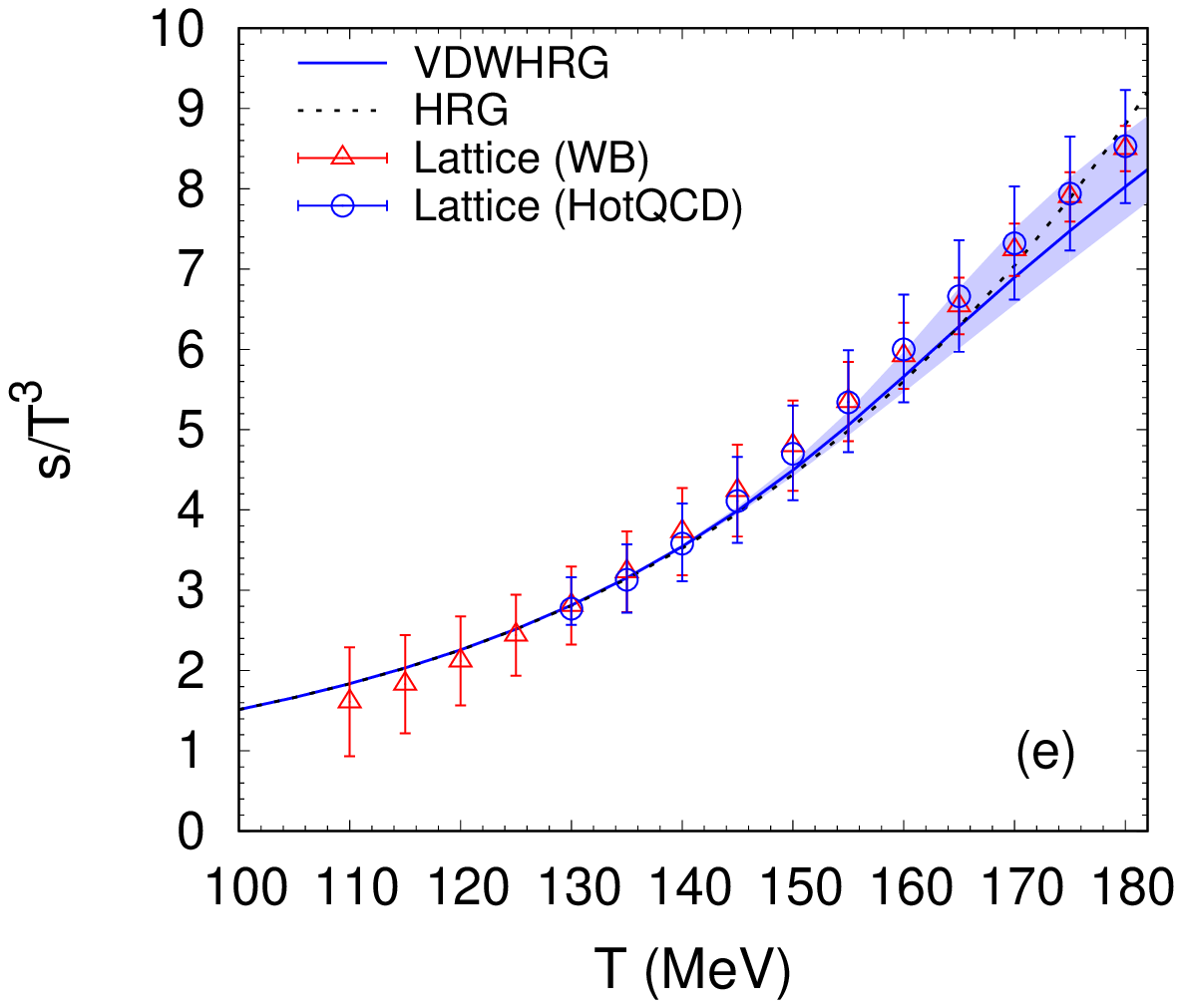}\label{sT3_T}
    \includegraphics[width=0.3\textwidth]{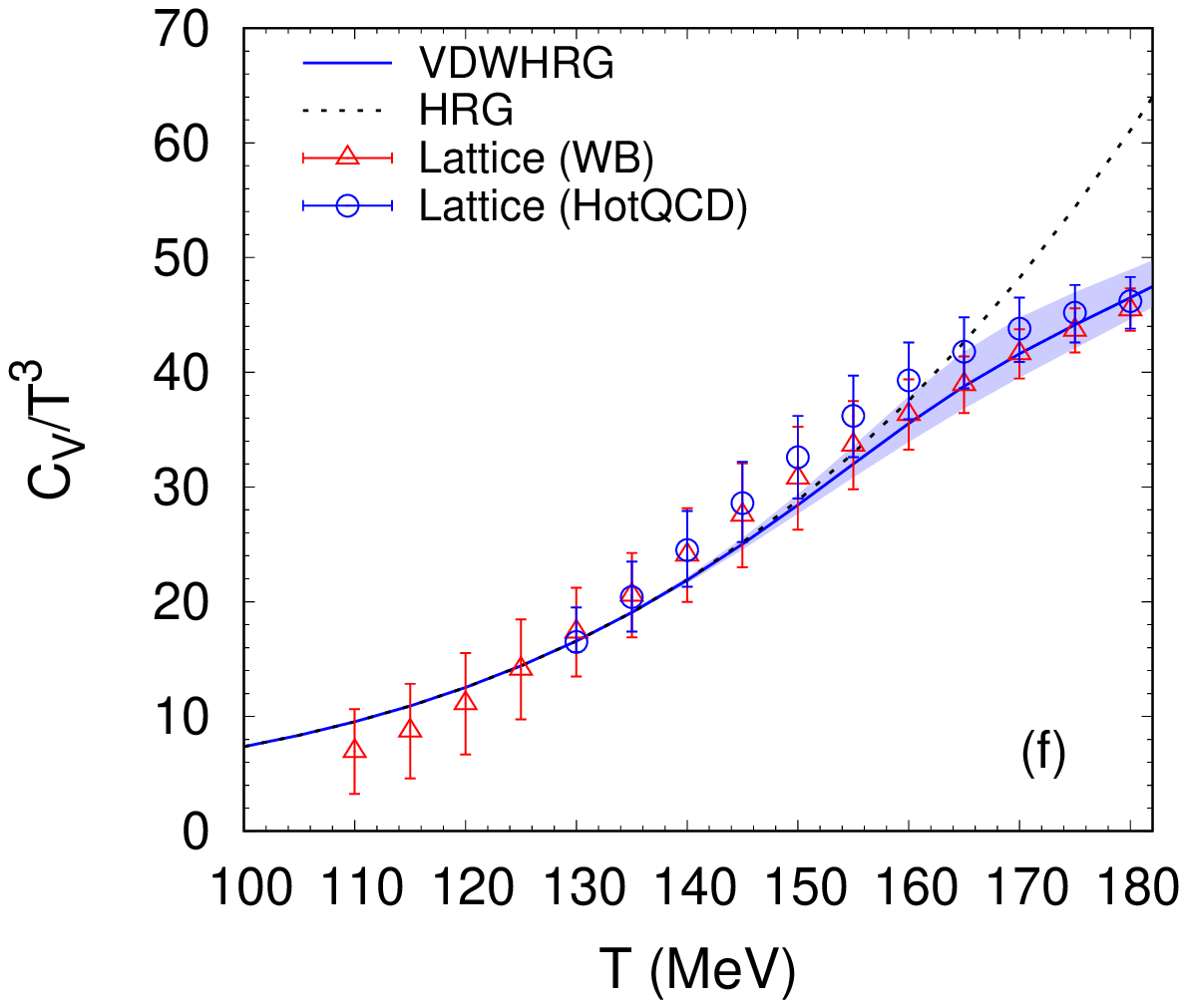}\label{CvT3_T}
     \vspace{0.7cm}
\caption{(Color online) The variation of different thermodynamical quantities 
with the temperature at $\mu = 0$. Blue lines show the results of VDWHRG model
using the parameters $a = 1250$ MeV fm$^3$ and $r = 0.7$ fm. 
Blue bands are due to the errors on the van der Waals
parameters in the VDWHRG model. 
The continuum extrapolated LQCD data are taken
from Refs. \cite{Borsanyi:2013bia} (WB) and \cite{Bazavov:2014pvz} (HotQCD).}
 \label{fig:fit}
\end{figure*}

\begin{figure*}[]
\centering
\includegraphics[width=0.3\textwidth]{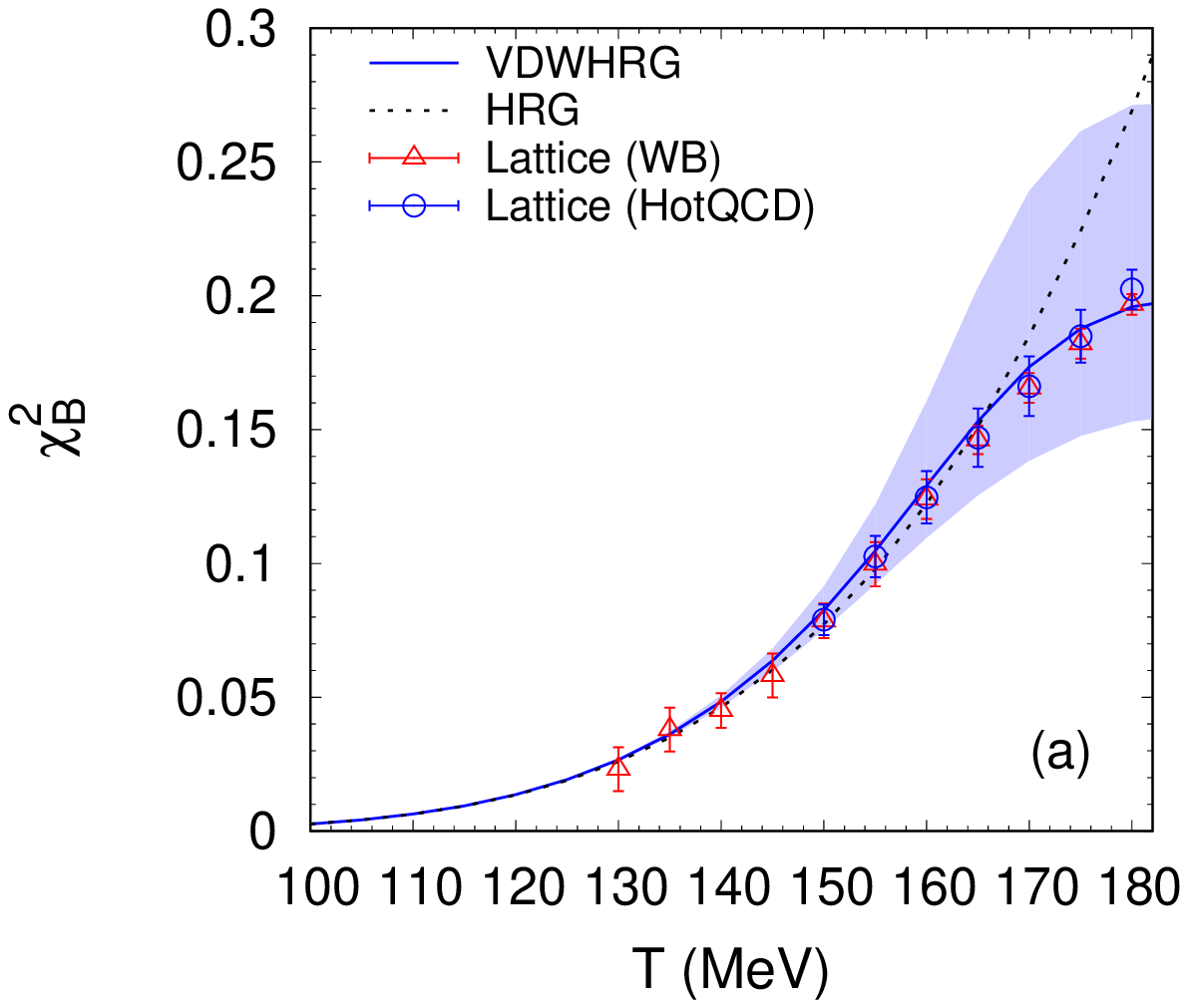}\label{chi2B_T}
 \includegraphics[width=0.3\textwidth]{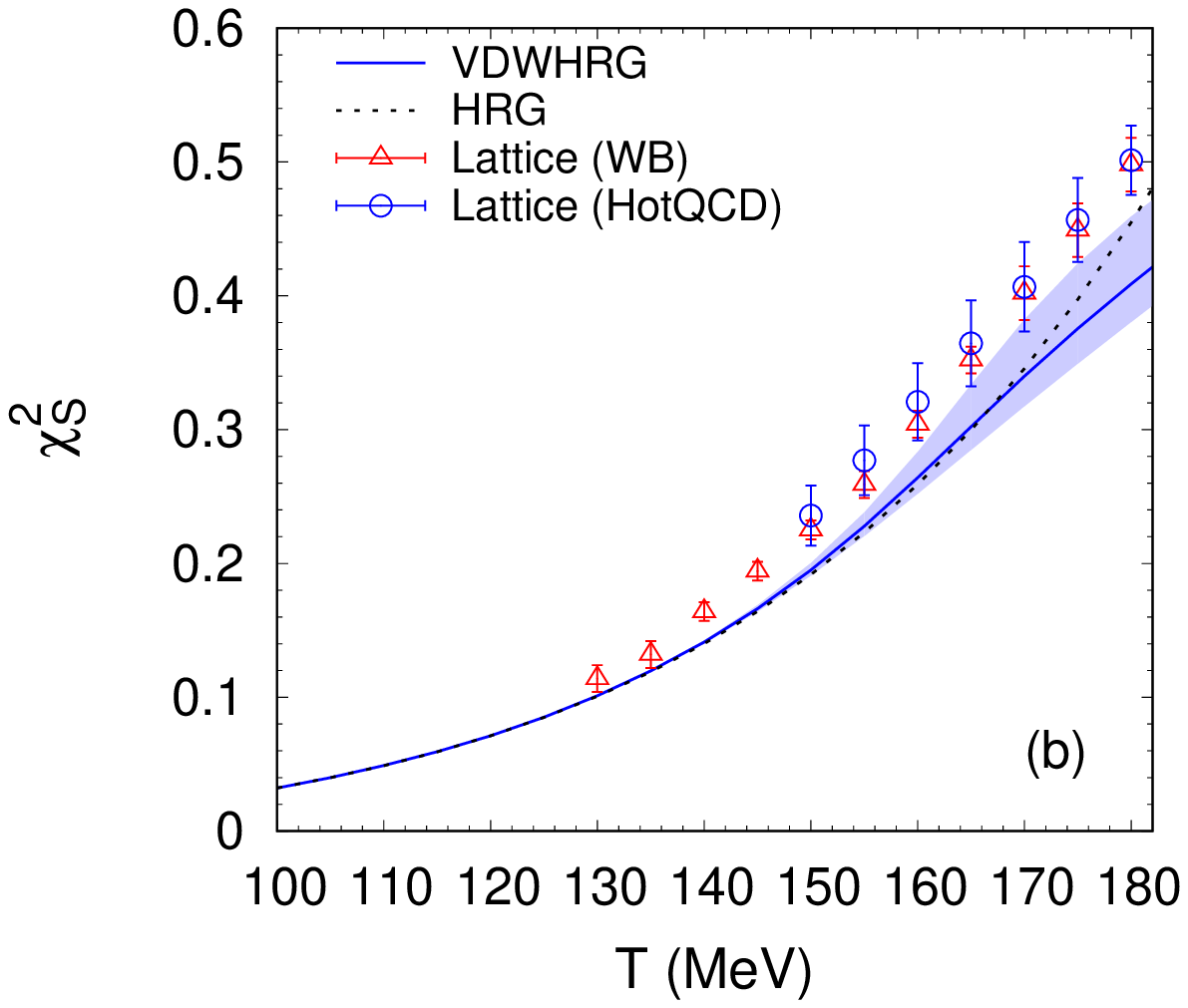}\label{chi2S_T}
  \includegraphics[width=0.3\textwidth]{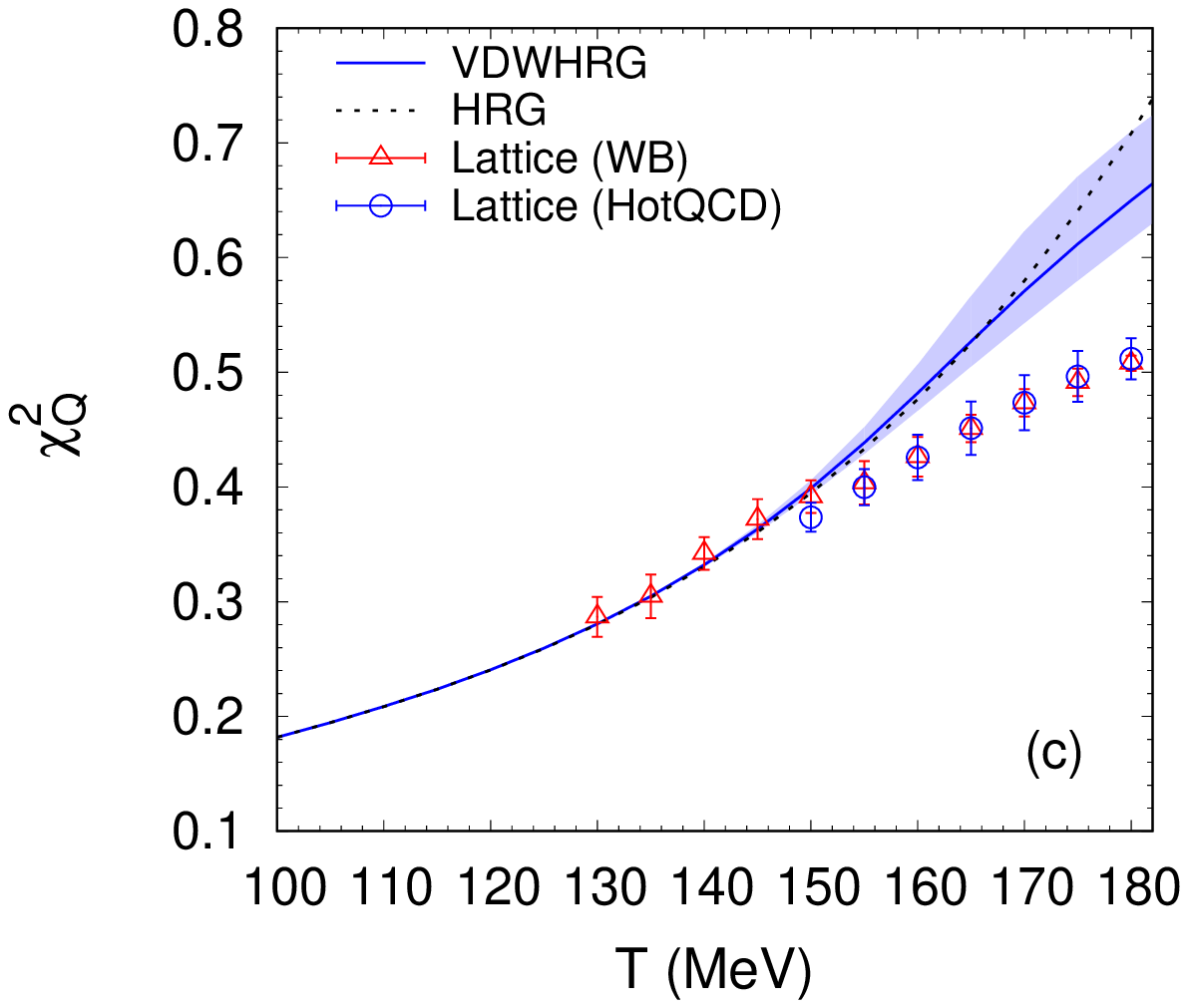}\label{chi2Q_T}
     \vspace{0.7cm}
\caption{(Color online) The variation of second order fluctuations of conserved charges with the temperature
at zero chemical potential. Blue lines show the results of VDWHRG model
using the parameters $a = 1250$ MeV fm$^3$ and $r = 0.7$ fm. Blue bands are due to the errors on the van der Waals
parameters in the VDWHRG model. The LQCD data are taken from 
Refs. \cite{Bellwied:2015lba,Bazavov:2012jq}.
}
 \label{fig:sus}
\end{figure*}

\section{\label{sec:result} Results}

In order to extract the van der Waals parameters $a$ and $r$ (or $b$) in VDWHRG model
that best describe the LQCD data at $\mu_B =0$, we 
use $\chi^2$ minimization technique where $\chi^2$ is defined as
\begin{equation}\label{eq:chi2}
 \chi^2 = \frac{1}{N} \sum_{i,j} \frac{(R_{i,j}^{LQCD} (T_j)-R_{i,j}^{model} (T_j))^2}{(\Delta_{i,j}^{LQCD} (T_j))^2},
\end{equation}
where $R_{i,j}^{model} (T_j)$ is the $i$th observable with 
$R_{i,j}^{LQCD} (T_j)$
and $\Delta_{i,j}^{LQCD} (T_j)$ are its values and errors
respectively at $j$th temperature calculated in LQCD, 
$N$ is the number of LQCD data points.
Here, we assume that van der Waals parameters $a$ and $r$ are independent of 
temperature and chemical potential.
Errors on the parameters are obtained by knowing their values at  $\chi^2_{min}+1$.
In this work we use latest continuum limit
LQCD data \cite{Borsanyi:2013bia,Bellwied:2015lba} of $p/T^4, \varepsilon/T^4, s/T^3, C_V/T^3$ and
$\chi^2_B$ at $\mu = 0$ within the temperature range $130 - 180$ MeV to calculate $\chi^2$ 
using Eq. \ref{eq:chi2}. 
\revision{
We assume that HRG model is valid up to $T = 180$ MeV because the transition 
at $\mu_B = 0$ is a crossover. Hence thermodynamic observables do
not exhibit sharp changes. LQCD results of quantities like $p/T^4, \varepsilon/T^4$ and $s/T^3$
have the smooth crossover at temperature range up to 180 MeV \cite{Bazavov:2012jq}.
Depending on the choice of order parameter the QCD crossover temperature ($T_c$) has a range of from
155 MeV to 175 MeV. For example LQCD calculation with chiral condensate gives 
$T_c = 155$ MeV \cite{Bazavov:2011nk}. However if one chooses 
strange quark number susceptibility the $T_c \sim$ 170 MeV \cite{Borsanyi:2010bp}.
Typical error including systematics and due to the choice of order parameter on is $T_c \sim$ 20 MeV.
}
Lowest temperature is taken as $T = 130$ MeV since the LQCD data of susceptibilities are not available
below $T = 130$ MeV in 
Ref. \cite{Bellwied:2015lba}. 
The best fit in terms of $\chi^2$ is 
achieved for parameter values of $a = 1250 \pm 150$ MeV fm$^3$ and $r = 0.7 \pm 0.05$ fm.
\revision{
Relatively smaller parameter values, $a = 329$ MeV fm$^3$ and $r = 0.59$ fm, were obtained
by \cite{Vovchenko:2015vxa}. With these parameters only a qualitative description
of LQCD data at $\mu =0$ is possible  ~\cite{Vovchenko:2016rkn} which we have already stated.}
In some previous works hardcore radius has been estimated in the EVHRG model.
% can comparition with the corresponding measurements.
In Ref. \cite{Yen:1997rv} hardcore radii of pion and other 
hadrons were obtained as $0.62$ fm and $0.8$ fm
respectively by fitting the experimental data of hadronic ratios at AGS and SPS energies.
While the value of the hardcore radius was estimated as $0.3$ fm in the 
Ref \cite{BraunMunzinger:1999qy} using the experimental data of hadronic 
ratios at SPS energies. 
Also in Refs. \cite{Andronic:2012ut,Bhattacharyya:2013oya} it was shown that the LQCD data of 
different thermodynamic quantities can be described in EVHRG model with the 
radius parameter between $0.2 -0.3$ fm.
Our present estimate of radius parameter is comparable to that of Ref. \cite{Yen:1997rv}.
However, it should be noted that in all those works 
\cite{Yen:1997rv,BraunMunzinger:1999qy,Andronic:2012ut,Bhattacharyya:2013oya}
only repulsive interaction was considered for 
all mesons and baryons and there was no attractive interaction.
\revision{To check the sensitivity of value of the parameters on the  temperature range
we have refitted the LQCD data up to $T = 165$ MeV (typical chemical freeze-out 
temperature from RHIC top energy) and found the new $a$ value to be 1210 MeV fm$^3$ 
which is within the uncertainty of the $a$ value obtained by fitting up to 
$T = 180$ MeV, i.e., 1250 MeV fm$^3 \pm 150$ MeV fm$^3$. There is no change in the 
value of $r$ parameter.}

Figure \ref{fig:fit} shows variation of $p/T^4, \varepsilon/T^4$, $(\varepsilon -3 p)/T^4$,
$c_s^2 = \partial p/ \partial \varepsilon$, $s/T^3, C_V/T^3$ and
$\chi^2_B$ with temperature at $\mu = 0$. Blue lines show the results of VDWHRG model
using the parameters $a = 1250$ MeV fm$^3$ and $r = 0.7$ fm. The bands are due to the errors on the parameters
$a$ and $r$.
Results of ideal HRG model along with the
LQCD data of the Wuppertal-Budapest (WB) Collaboration \cite{Borsanyi:2013bia} and 
the Hot QCD Collaboration
\cite{Bazavov:2014pvz} are also shown in this figure.
Our estimations of all these observables in the VDWHRG model
are in good agreement with LQCD calculations in
the temperature range studied. Compared to ideal HRG model, improvement
of the results in VDWHRG model is observed
which indicates the interacting nature of baryons especially at high temperature region.
Among all these observables, behavior of $c_s^2$ is most interesting in VDWHRG model.
The $c_s^2$ is a quantity that is sensitive to the phase transition effect.
While in ideal HRG model $c_s^2$
decreases with increasing temperature, in VDWHRG model it shows a minimum near $T = 150$ MeV
which is consistent with the LQCD data.
The minimum of the $c_s^2$ is known as the softest point where the
expansion of the system slows down. 
As a result the system spends a longer time in this temperature range
which may be a crucial indicator of the quark-hadron transition of the system
observed in heavy ion collisions \cite{Hung:1994eq, Bazavov:2014pvz}.

In Fig. \ref{fig:sus} temperature dependence of
second order fluctuations of different conserved
charges at zero chemical potential have been shown. 
One can see that the qualitative behaviors of all these 
fluctuations in VDWHRG model are similar to  the LQCD data
at high temperature which are different from ideal HRG model
where all the quantities increase rapidly with increasing temperature.
Not only that, the $\chi^2_B$
and $\chi^2_S$ obtained from the VDWHRG model match quantitatively with the LQCD data.
However, for $\chi^2_Q$, which is dominated by the non-interacting mesons, hence
VDWHRG model overestimates the LQCD data.

\begin{figure*}[]
\centering
   \includegraphics[width=0.3\textwidth]{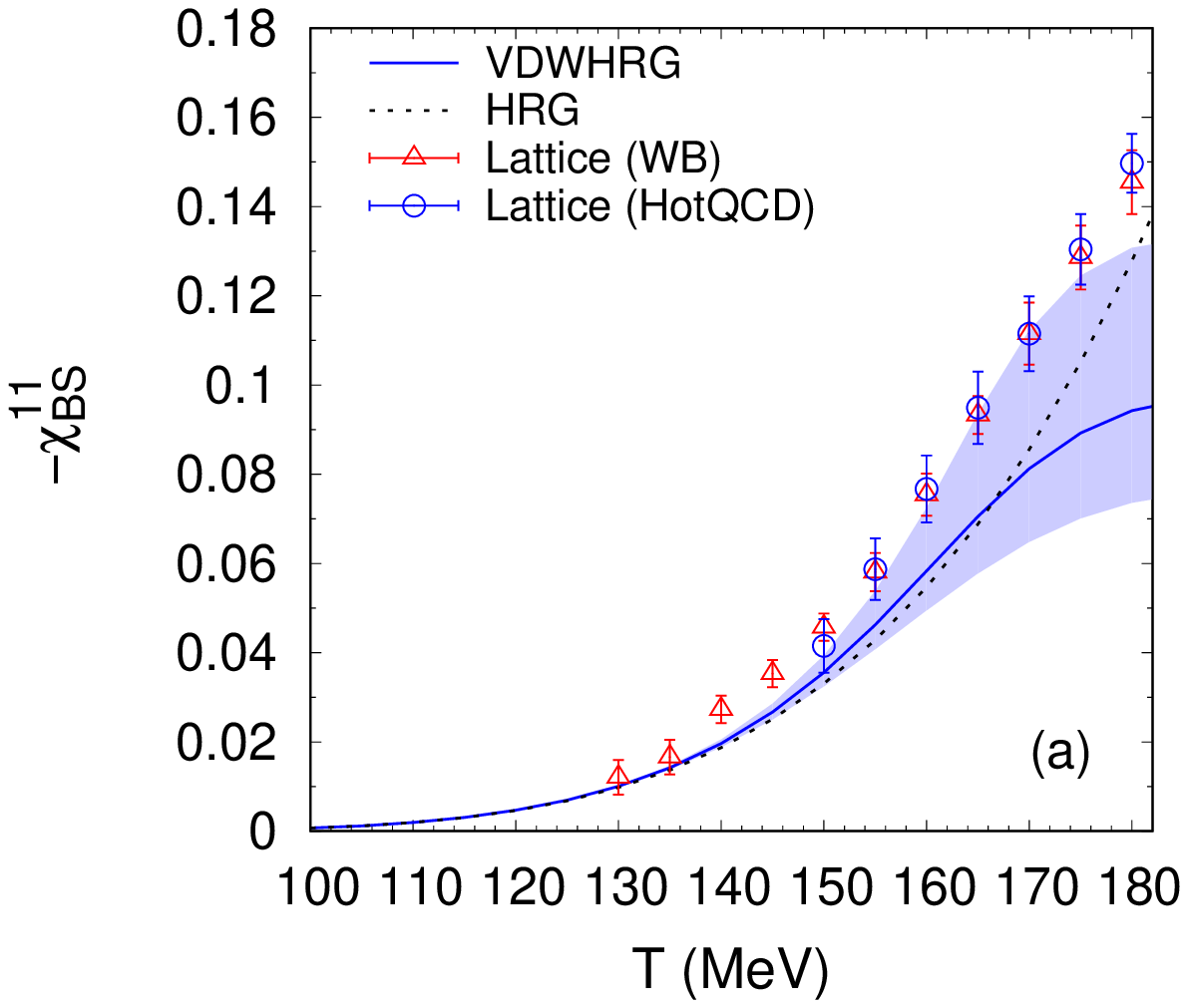}\label{chiBS11_T}
    \includegraphics[width=0.3\textwidth]{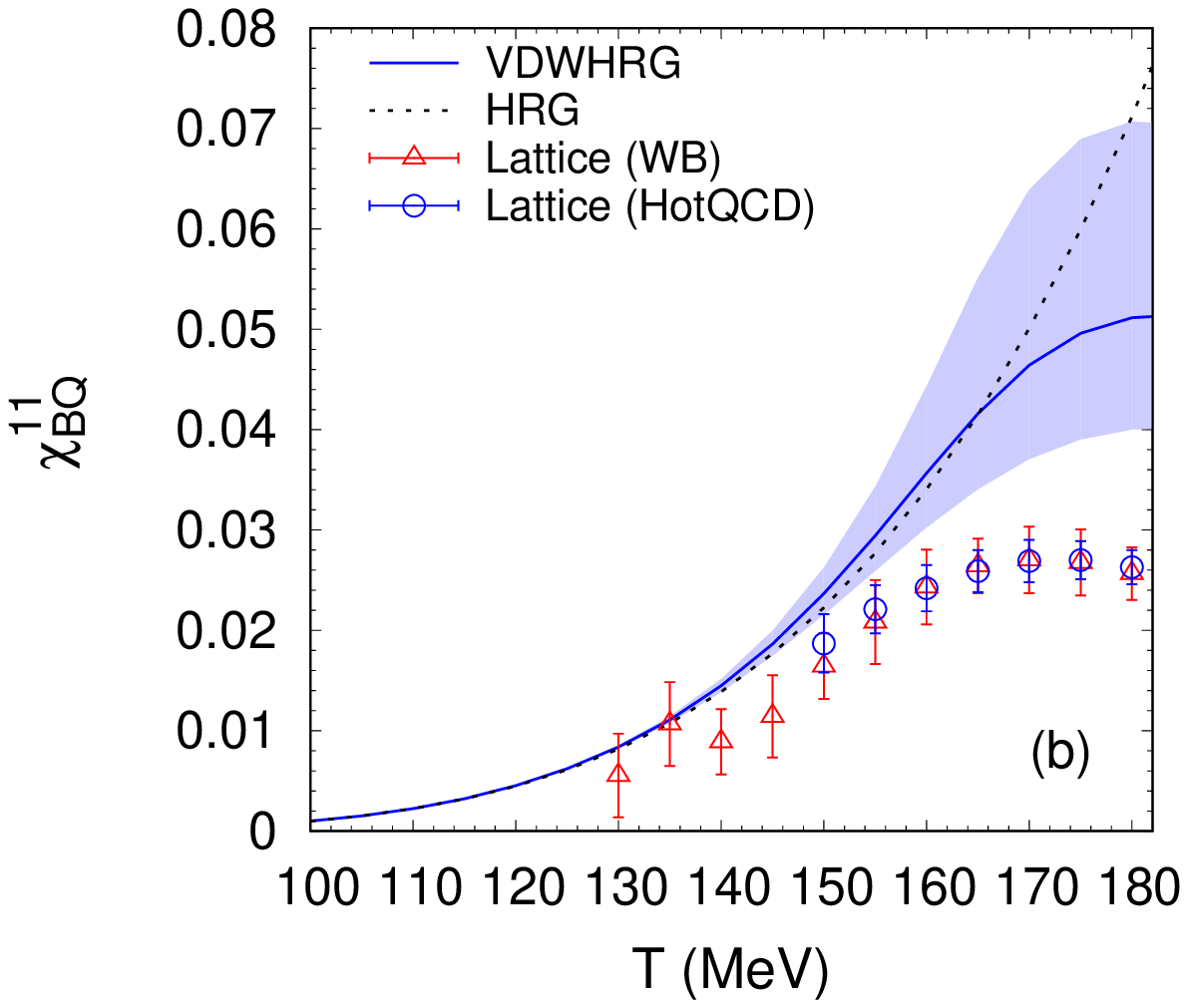}\label{chiBQ11_T}
     \includegraphics[width=0.3\textwidth]{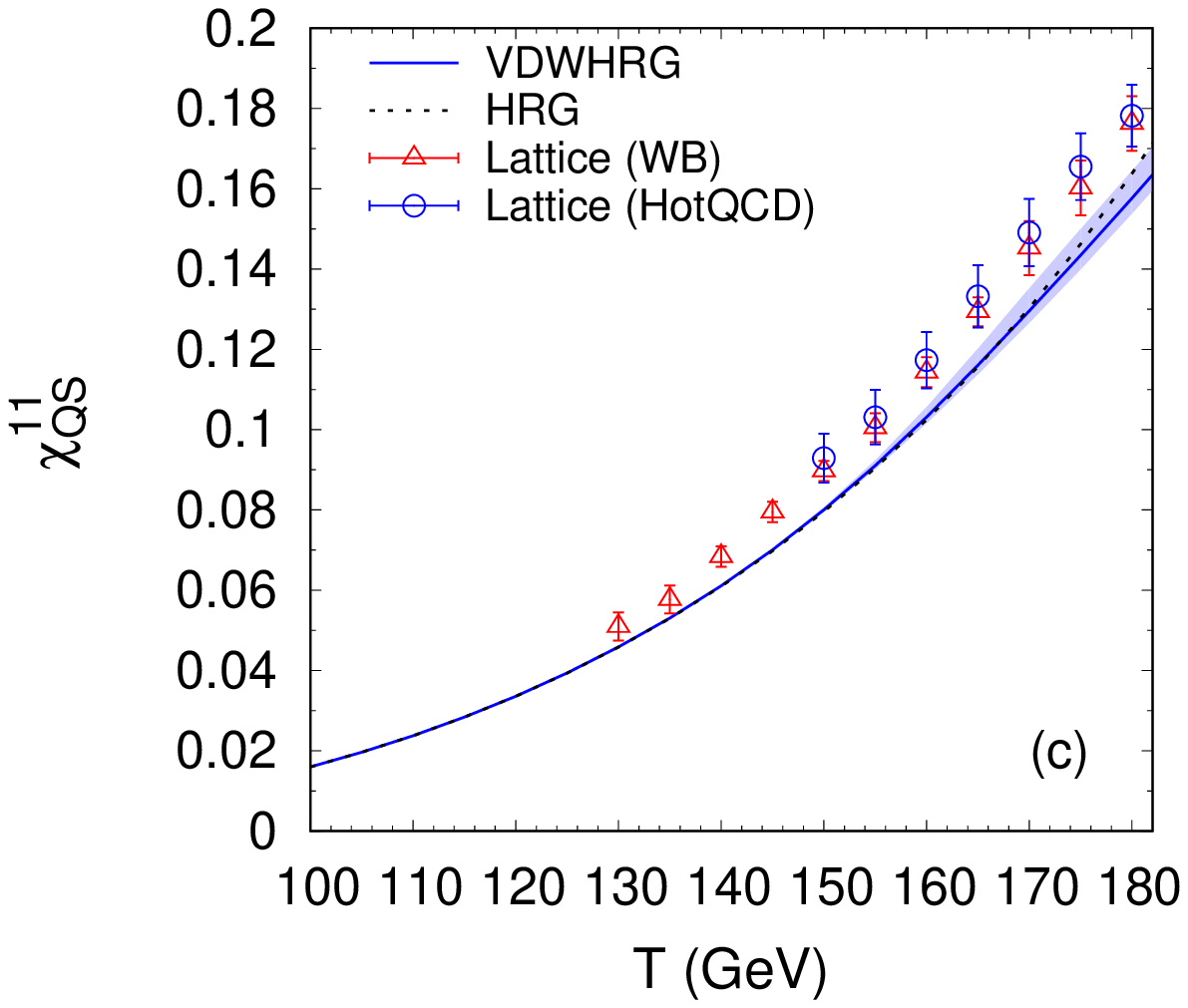}\label{chiQS11_T}
     \vspace{0.7cm}
\caption{(Color online) The variation of correlations between conserved charges with
the temperature at zero chemical potential. Blue lines show the results of VDWHRG model
using the parameters $a = 1250$ MeV fm$^3$ and $r = 0.7$ fm. Blue bands are due to the errors on the van der Waals
parameters in the VDWHRG model. The LQCD data are taken from 
Refs. \cite{Bellwied:2015lba,Bazavov:2012jq}.}
 \label{fig:correction}
\end{figure*}

\begin{figure}[]
\centering
\includegraphics[width=0.4\textwidth]{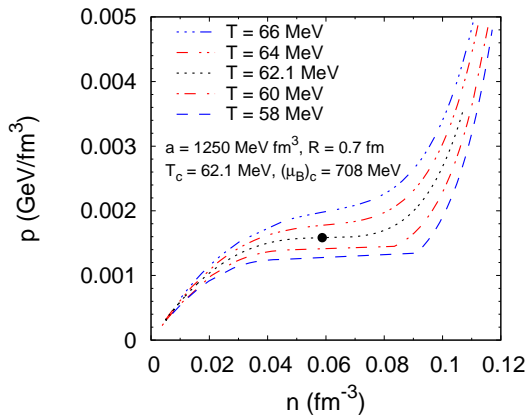}
\vspace{0.7cm}
\caption{(Color online) The variation of pressure with the number density of the 
hadronic medium at different temperature. The black dot indicates the critical point.}
 \label{fig:p_n}
\end{figure}

\begin{figure}[]
\centering
\includegraphics[width=0.4\textwidth]{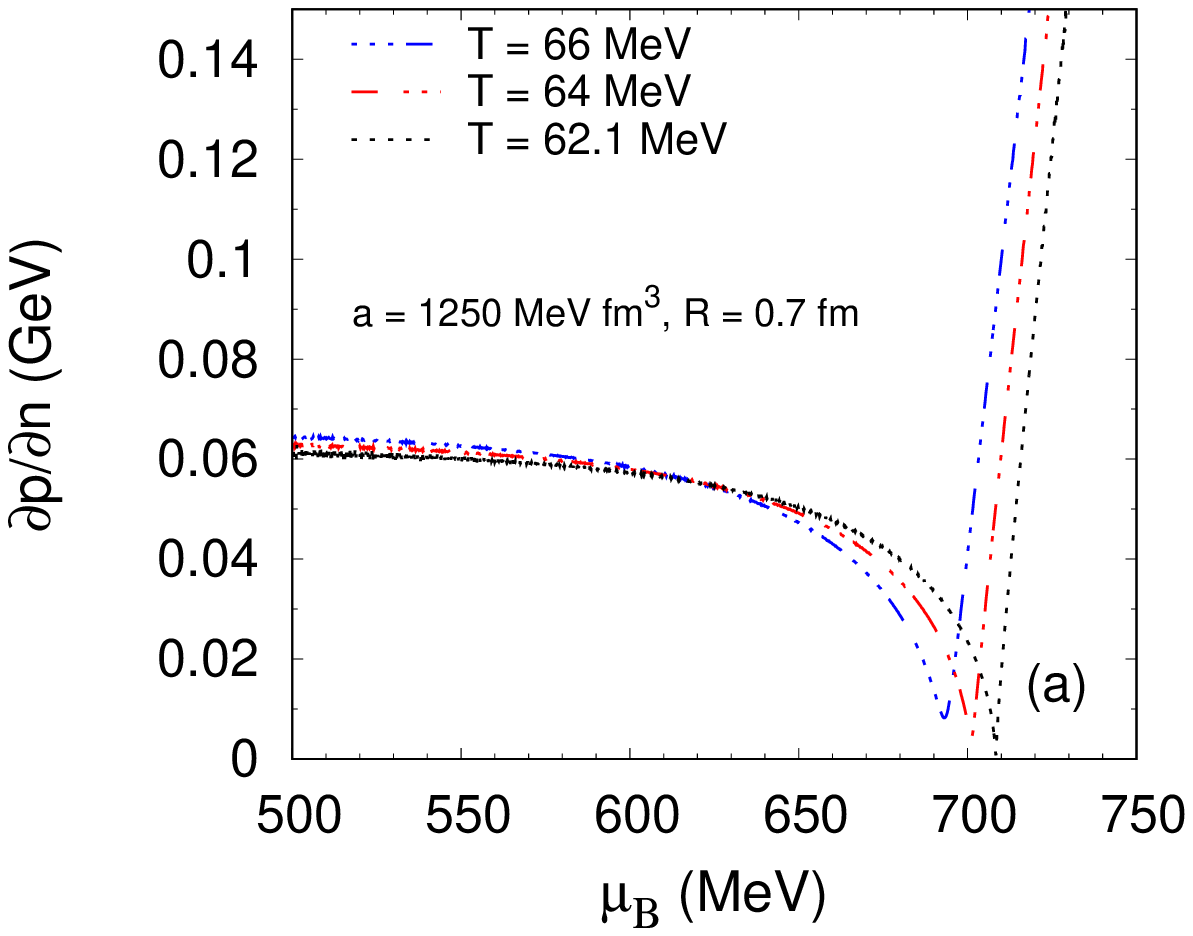}
\includegraphics[width=0.4\textwidth]{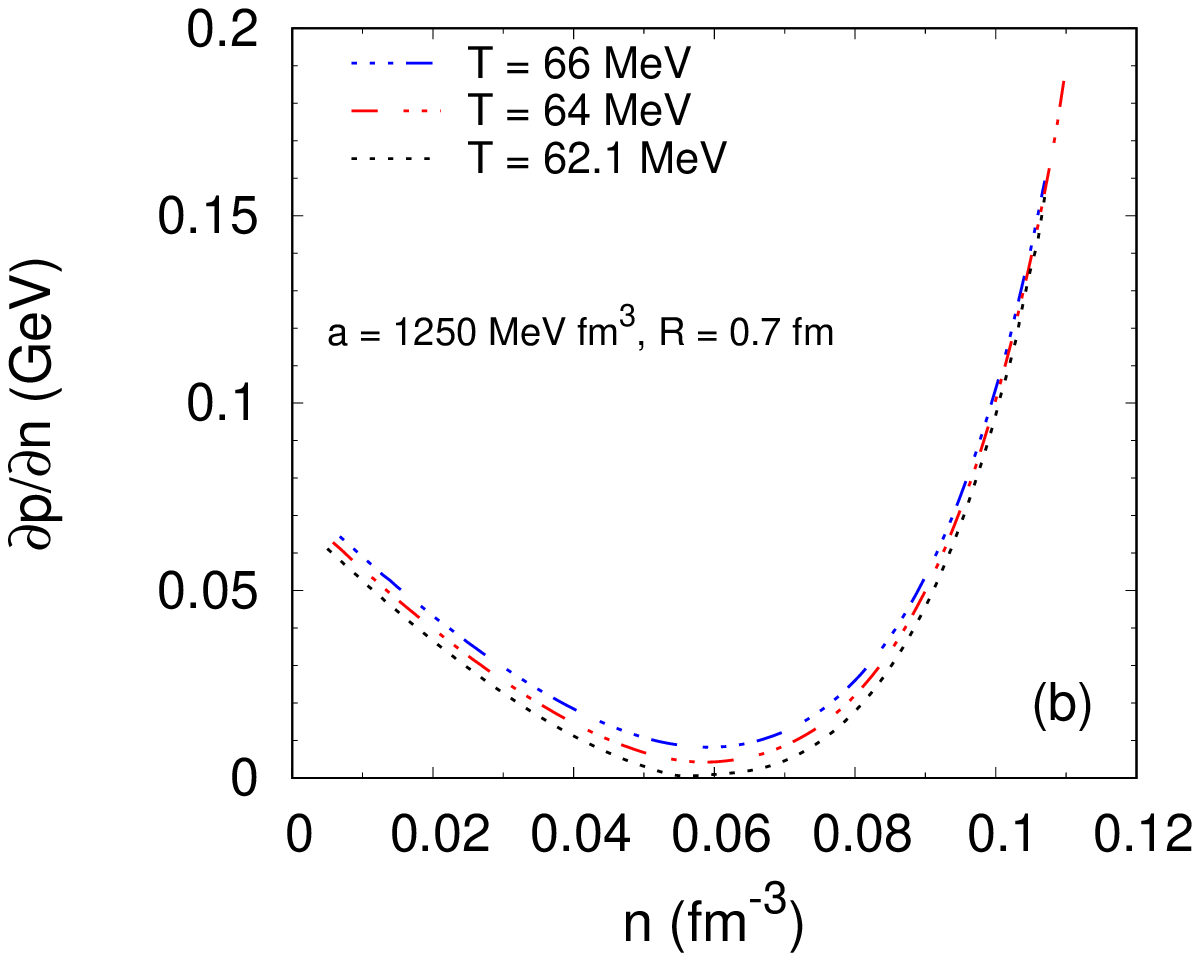}
\vspace{0.7cm}
\caption{(Color online) Variations of $(\partial p/\partial n)_T$
with respect to $\mu_B$ and $n$ respectively.}
 \label{fig:dp_dn}
\end{figure}

\begin{figure}[]
\centering
\includegraphics[width=0.4\textwidth]{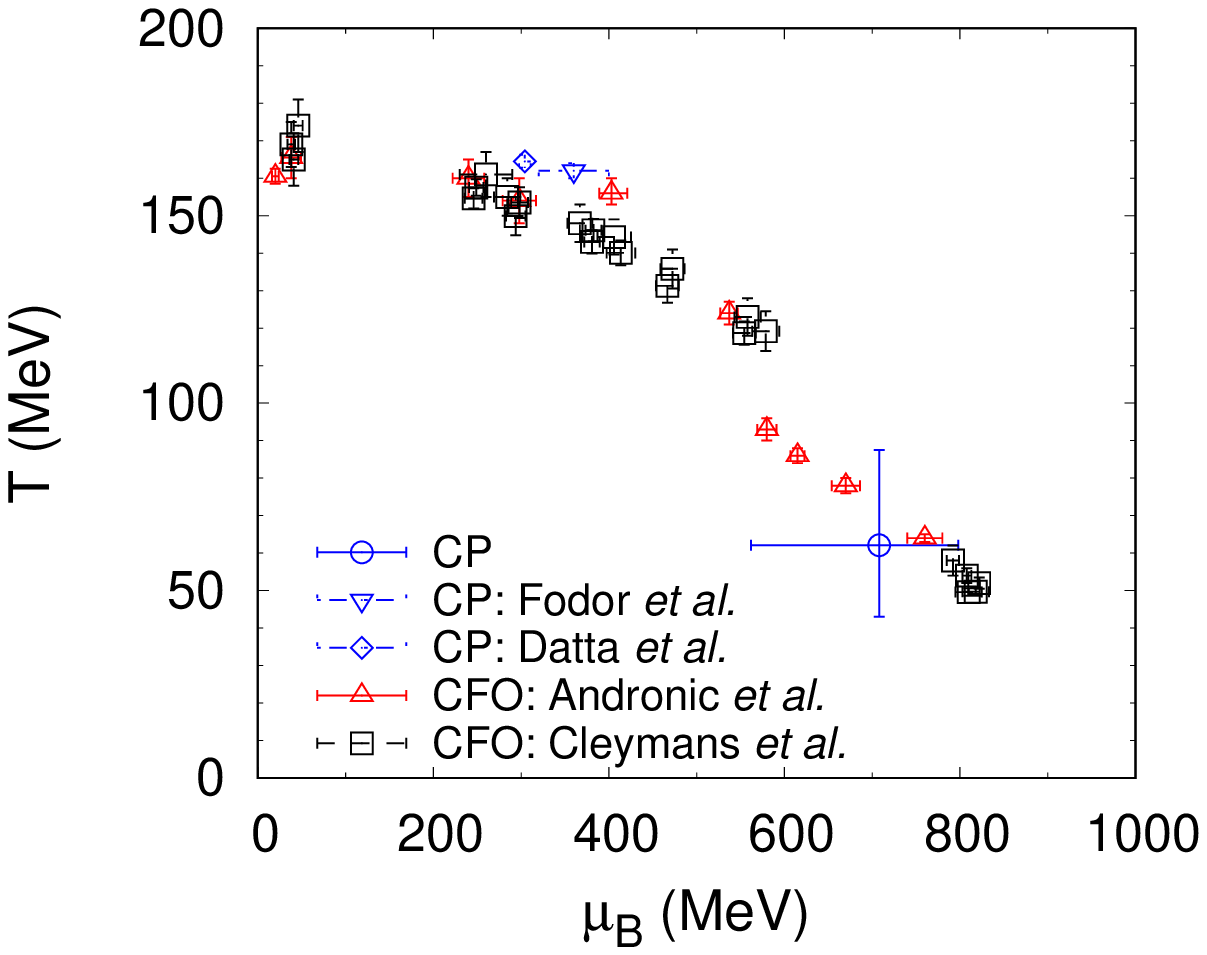}
\vspace{0.7cm}
\caption{(Color online)
The critical point (CP) of the liquid-gas transition of the present work in the QCD phase diagram.
Critical point calculated in lattice are taken from Ref. \cite{Fodor:2004nz} (Fodor et al.)
and Ref. \cite{Datta:2016ukp} (Datta et al.).
Chemical freeze-out (CFO) parameters shown in this figure are taken from 
Ref. ~\cite{Andronic:2005yp} (Andronic et al.) and
Ref. ~\cite{Cleymans:2005xv} (Cleymans et al.).}
 \label{fig:phase}
\end{figure}

Figure \ref{fig:correction} shows correlations among conserved charges. 
Magnitudes of the $\chi^{11}_{BS}$ and $\chi^{11}_{QS}$ increase with
increasing temperature and at a very high temperature they are expected to
reach at $1/3$, the value at Stefan-Boltzmann limit.
$\chi^{11}_{BS}$ and $\chi^{11}_{QS}$ calculated in VDWHRG model are close to the 
LQCD data in the temperature range studied.
On the other hand, the LQCD data of $\chi^{11}_{BQ}$ shows
a hump around $T = 170$ MeV which indicate the crossover transition in this region.
Almost a similar qualitative behavior is observed for $\chi^{11}_{BQ}$ in VDWHRG model as well
although the model values overestimate the LQCD data.

The HRG model does not have QGP phase but with attractive and repulsive
interaction for baryons VDWHRG model explains LQCD data which have QGP phase.
So if interactions are the sole driving force behind the physics
of phase transitions one expects a similar phase transition effect in VDWHRG model.
Figure \ref{fig:p_n} shows variation of pressure with number density
at a fixed temperature in VDWHRG model.
The parameters $a$ and $r$ are fixed from the best fit values of the 
VDWHRG model to LQCD data at $\mu_B =0$.
\revision{For simplicity, we assume nature of the interaction is similar to both non zero
and zero $\mu_B$ regions of the phase diagram.}
We observe the value of critical temperature to be $T= 62.1$ MeV. 
Below this temperature the number
density changes discontinuously which resembles a hadron-liquid first order 
phase transition.
The picture will be more clear in
Fig. \ref{fig:dp_dn}, where we show variations of $(\partial p/\partial n)_T$ 
with respect to $\mu_B$ and $n$ respectively. One can see that at
$T = 62.1$ MeV and $\mu_B = 708$ MeV, $(\partial p/\partial n)$ becomes zero
and above $T = 62.1$ MeV, $(\partial p/\partial n)$ is always greater than zero.
\revision{
Since we have used van der Waals interaction it is expected that the phase 
transition which we observed is a liquid-gas phase transition and the critical
point ($T= 62.1^{+25.4}_{-19.1}$ MeV, $\mu_B = 708^{+90}_{-146}$ MeV)
so obtained is that of a liquid-gas transition.}
Errors on the critical point is due to the uncertainties on the parameters $a$ and $r$.
A similar result of critical point with $T = 89$ MeV and $\mu_B = 724$ MeV
is also obtained by using the holographic gauge/gravity correspondence
to map baryon number fluctuations in QCD to the charge
fluctuations of holographic black holes \cite{Critelli:2017oub}.
% However the properties of the liquid are unknown in this study.
% Baryon chemical potential corresponding to the critical point is $\mu_B = 708^{+90}_{-146}$ MeV.

\revision{In the Fig. 6 we have plotted a collection of ($T,\mu_B$) points to make a
comparison of (i) liquid-gas CP from our present analysis (ii) CP from LQCD (iii)
chemical freeze-out parameters from heavy ion collision experiments.}
Blue circular point in Fig. \ref{fig:phase} 
shows the critical point ($T= 62.1^{+25.4}_{-19.1}$ MeV, $\mu_B = 708^{+90}_{-146}$ MeV) 
of the liquid-gas transition as estimated within the current model calculations
in the QCD phase diagram. 
Critical points calculated
in lattice \cite{Fodor:2004nz, Datta:2016ukp}
and the chemical freeze-out parameters obtained 
by different groups ~\cite{Andronic:2005yp, Cleymans:2005xv}
at various energies are also shown in this plot.

\section{\label{sec:conclusions}Summary}
To summarize, we have used LQCD data of $p/T^4, \varepsilon/T^4, s/T^3, C_V/T^3$ and
$\chi^2_B$ at $\mu = 0$ to extract the van der Waals parameters in the VDWHRG model.
We assume that baryons are interacting whereas mesons are non-interacting.
We get $a = 1250 \pm 150$ MeV fm$^3$ and $r = 0.7 \pm 0.05$ fm in our present work
which best describes the LQCD data at $\mu =0$ within the temperature range $130 - 180$ MeV.
The values of the VDWHRG model parameters are obtained using a chi-square minimization
procedure.
With these parameters which explains the QCD matter simulated by lattice, we observe a
phase transition in VDWHRG model
at large  potential with a critical point in the $(T, \mu_B)$ phase diagram
at $T = 62.1$ MeV and $\mu_B = 708$ MeV.
Our result of critical point is comparable with that of Ref. \cite{Critelli:2017oub}
where the critical point is obtained by using the holographic gauge/gravity correspondence
to map baryon number fluctuations in QCD to the charge
fluctuations of holographic black holes.
Several improvements in the future can be carried out to our present idea and work.
One of them includes incorporating the mesonic interaction of the system. 
Another is to incorporate other missing
resonances in the hadronic spectrum \cite{Bazavov:2014xya}.

\section*{Acknowledgements}
We are thankful to Sourendu Gupta, Victor Roy and Nu Xu for carefully
reading the manuscript and for their valuable comments.
We thank Paolo Alba for suggesting useful references related to the work.
BM acknowledges financial support from J C Bose National Fellowship of DST, Government
of India. SS acknowledges financial support from DAE-SRC, Government of India.
% The authors appreciate the valuable comments of Nu Xu.
% , Sourendu Gupta and Victor Roy.

\end{document}